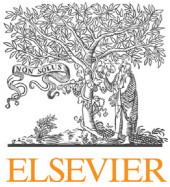

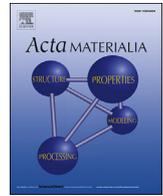

Full length article

# High temperature internal friction in a Ti−46Al−1Mo−0.2Si intermetallic, comparison with creep behaviour

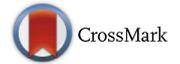

M. Castillo-Rodríguez [a, d], M.L. Nó [b], J.A. Jiménez [c], O.A. Ruano [c], J. San Juan [a, *]

[a] Dpto Física de la Materia Condensada, Facultad de Ciencia y Tecnología, Universidad del País Vasco, Aptdo 644, 48080 Bilbao, Spain
[b] Dpto Física aplicada II, Facultad de Ciencia y Tecnología, Universidad del País Vasco, Aptdo 644, 48080 Bilbao, Spain
[c] Dpto Metalurgia Física, Centro Nacional de Investigaciones Metalúrgicas (CENIM-CSIC), 28040 Madrid, Spain
[d] Instituto de Ciencia de Materiales de Sevilla, CSIC-Universidad de Sevilla, Avda. Américo Vespucio 49, 41092 Sevilla, Spain

## ABSTRACT

Advanced γ-TiAl based intermetallics Mo-bearing have been developed to obtain the fine-grained microstructure required for superplastic deformation to be used during further processing. In the present work we have studied an alloy of Ti−46.8Al−1Mo−0.2Si (at%) with two different microstructures, as-cast material with a coarse grain size above 300 μm, and the hot extruded material exhibiting a grain size smaller than 20 μm. We have used a mechanical spectrometer especially developed for high temperature internal friction measurements to study the defect mobility processes taking place at high temperature. The internal friction spectra at different frequencies has been studied and analyzed up to 1360 K in order to characterize the relaxation processes appearing in this temperature range. A relaxation peak, with a maximum in between 900 K and 1080 K, depending on the oscillating frequency, has been attributed to Ti-atoms diffusion by the stress-induced reorientation of Al−V$_{Ti}$−Al elastic dipoles. The high temperature background in both microstructural states, as-cast and extruded, has been analyzed, measuring the apparent activation parameters, in particular the apparent energies of E$_{cast}$(IF) = 4.4 ± 0.05 eV and E$_{ext}$(IF) = 4.75 ± 0.05 eV respectively. These results have been compared to those obtained on the same materials by creep deformation. We may conclude that the activation parameters obtained by internal friction analysis, are consistent with the ones measured by creep. Furthermore, the analysis of the high temperature background allows establish the difference on creep resistance for both microstructural states.

© 2015 Acta Materialia Inc. Published by Elsevier Ltd. This is an open access article under the CC BY-NC-ND license (http://creativecommons.org/licenses/by-nc-nd/4.0/).

## 1. Introduction

γ-TiAl-based alloys are one of the most important materials targeted for industrial applications in particular in aerospace and automotive engines. Among its amazing properties, a low specific weight (3.8−4.1 g/cm$^3$), good oxidation and burn resistance (up to 800 °C) together with a high elastic stiffness and enhanced high temperature strength are remarkable [1−5]. These exceptional properties made these alloys potentially attractive for applications under high thermal and mechanical load [6−8]. In addition those TiAl-based alloys submitted to an appropriate microstructural refinement may even exhibit super-plasticity [9−11], which from a technological point of view would notably improve the potential applications of the material. After the first generation of binary γ-TiAl alloys, a second generation of ternary alloys with several minor alloying elements and grain refiners was developed [12,13]. Then a third generation with a higher Nb content and B as grain refiner, called TNB, was designed to increase room-temperature ductility and high-temperature creep resistance [12,14]. Finally a fourth generation, called TNM, containing both Nb and Mo in order to stabilize the β phase, allowing a near conventional processing and creep-resistance for long-term service up to 1025 K [15], is being developed. Recent overviews on the development of γ-TiAl alloys can be found in Refs. [14,16].

The research on γ-TiAl has undergone a renewed interest since the announcement in 2010 of the introduction of cast blades, made of second generation γ-TiAl, in the low-pressure turbine of the General Electric engine [17], which at present is equipping the Dream Liner 787 from Boeing [18].

One of the key points to improve the new generation of γ-TiAl is to acquire a deep comprehension of the microscopic mechanisms controlling the high temperature creep deformation. In the last decade a new approach, by mechanical spectroscopy, is being used to study the mobility of defects at high temperature in structural intermetallics as FeAl [19–22] and TiAl [23–27]. A relationship between the internal friction high temperature background (HTB), measured by mechanical spectroscopy, and the high temperature creep behaviour in intermetallics has been proposed in several works [21,22,24,26] being still a matter of discussion.

* Corresponding author.
E-mail address: jose.sanjuan@ehu.es (J. San Juan).

In the present work we approach the study of the defects mobility on a prototype alloy with 1% Mo, which could be considered as belonging to the second alloy's generation, but with an added interest for the development of the fourth alloy's generation. This alloy has been also selected because in previous studies [28,29] its creep behaviour was studied and the activation enthalpies for two different microstructural conditions were determined through standard creep tests. In particular, the Ti−46Al−1Mo−0.2Si alloy in the as-cast and the hot extruded conditions show different creep properties but the activation energies for creep were very similar, 395 kJ/mol (4.1 eV) and 420 kJ/mol (4.35 eV) (1 eV = 96,2 kJ/mol) respectively [29]. These activation energies are similar to that for aluminium diffusion in TiAl. The creep behaviour, therefore, was related to a dislocation creep mechanism controlled by lattice diffusion of the slowest moving specie.

Independently of the particular deformation mechanism controlling creep, the internal friction high temperature background (HTB), measured by mechanical spectroscopy, may contribute to the determination of the atom mobility during deformation of a given material and, in turn, to determination of the temperature dependence of creep rate. This could be of technological interest, because creep tests are time consuming and usually require relatively large samples. In addition, mechanical spectroscopy is a non-destructive technique, use small samples and many experiments can be performed under different conditions in a reasonable time. So both techniques could be used collaboratively if their results could be reliably comparable.

## 2. Experimental details

### 2.1. Materials and microstructural characterization

Mo-alloyed two-phase γ/α TiAl material of composition (in at.%) Ti−46.8Al−1Mo−0.19Si was produced in 50 kg ingots by vacuum arc melting [28]. As-cast ingots were refined by hot extrusion in the α + γ field at 1300 °C. The reduction in area of the extruded rods was 7:1. The microstructure features, phases, grain size, morphology … were studied using a field emission SEM microscope (JEOL JSM-7000F) operating at 20 KV and 10 KV using backscattered electrons (BSE). Prior to observation, the samples were polished with diamond paste of grain size down to 1 μm, and subsequently with a colloidal silica suspension.

Fig. 1 shows SEM micrographs of the as-cast material. Its microstructure is composed of coarse lamellar grains bigger than 300 μm in size, Fig. 1a, formed by a quasi-eutectoid solid-state reaction. Inside the grains, coarse alternated γ-TiAl and $\alpha_2$-$Ti_3Al$ lamellae are observed, Fig. 1b, as well as some β phase at the grain boundaries and triple points, Fig. 1c. The identification of the phases is indicated in Fig. 1c and a complete description is included as Supplementary material. However, extruded material exhibits a finer and not completely homogeneous microstructure composed by different extrusion bands, Fig. 2a. On the one hand we observe bands of grains up to ~10 μm in size, Fig. 2b, but in general the extrusion bands are formed by α and γ grains < 5 μm in size, Fig. 2c. On the other hand, an appreciable amount of Mo-rich particles of β phase are scattered around the grains, Fig. 2c and d. The identification of the phases is indicated in Fig. 2d and a complete description is included as Supplementary material.

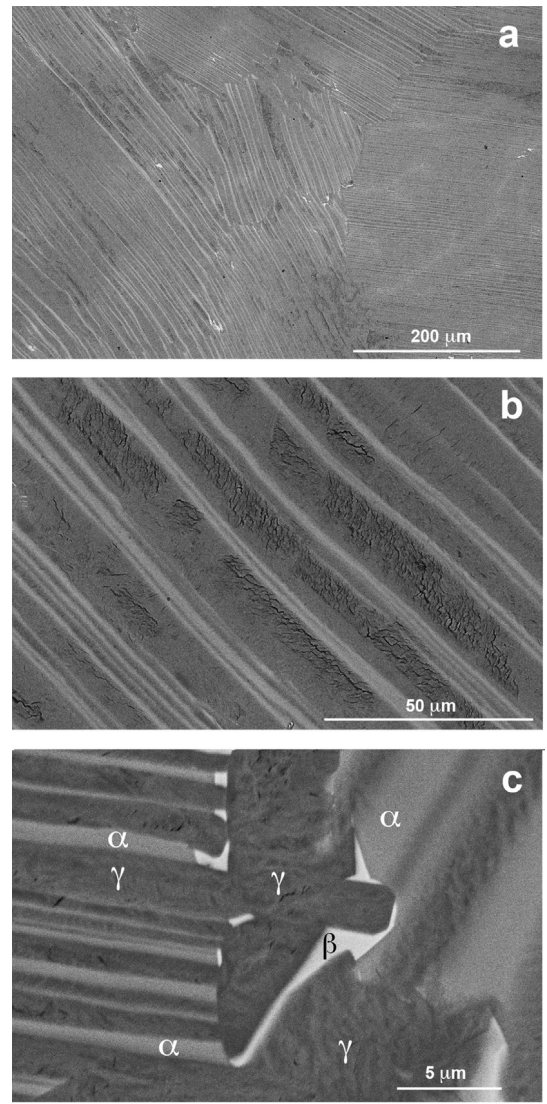

**Fig. 1.** (a) TEM-micrograph of the as-cast TiAl based alloy. (b) An alternation of γ-TiAl and $\alpha_2$-$Ti_3Al$ lamellae is observed inside the coarse lamellar grains. (c) Particles of β phase (white) are observed at grain boundaries and triple points.

### 2.2. Mechanical spectroscopy

The microscopic processes associated to the mobility of defects at high temperature have been investigated by mechanical spectroscopy. Samples were cut as parallelepipeds of approximate dimensions $50 \times 5 \times 1$ mm³, and lateral faces were polished with a diamond paste of grain size down to 1 μm. In this work the internal friction of both, the as-cast and extruded samples have been measured for various oscillating stress frequencies between 0.03 and 10 Hz in the temperature range from 600 to 1380 K.

Mechanical spectroscopy measurements have been carried out with an inverted torsion pendulum working in sub-resonant condition, measuring the internal friction $Q^{-1}$ through the delayed phase angle φ between the applied oscillating stress and the resulting oscillating strain. The internal friction can be measured in two different working modes: (a) as a function of temperature (300 K−1800 K) at imposed frequency, and (b) as a function of frequency ($10^{-3}$ Hz–10 Hz) in isothermal conditions. Detailed information about this equipment can be found elsewhere [30,31].

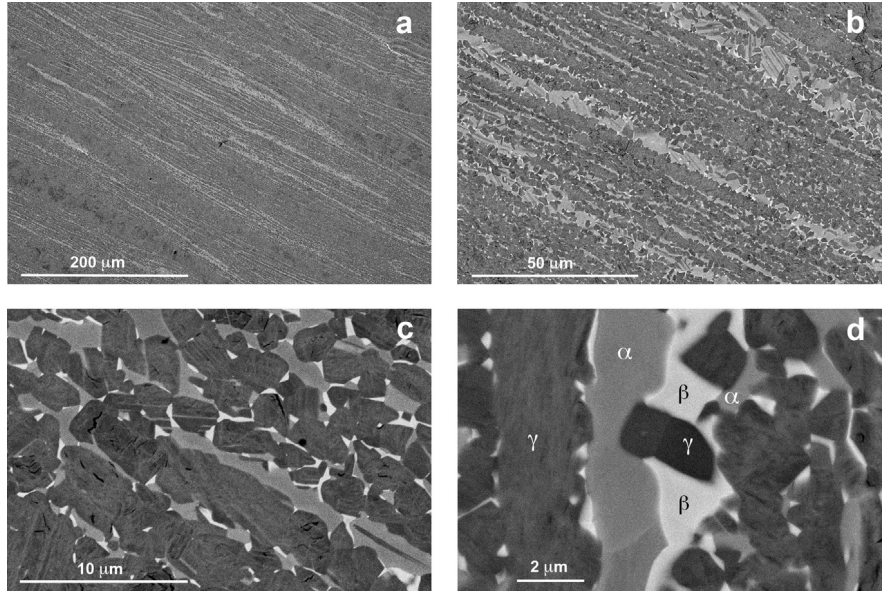

**Fig. 2.** (a) General view of the extruded material showing a non complete homogeneous microstructure; (b) bands of α₂ and γ grains with up to ~10 μm in size (c) bands of finer grains (<5 μm) of α₂ and γ phases with an appreciable amount of small particles of β phase. (d) Identification of the three phases.

In the present work we are facing the analysis of internal friction spectra as the ones of Fig. 3, showing a relaxation peak P1 at about 1000 K superimposed to a high temperature background (HTB) growing very fast above 1100 K.

The internal friction for a relaxation peak follows the expression:

$$Q^{-1} = \tan\phi(\omega) = \overline{\Delta}\cdot\frac{\omega\overline{\tau}}{1+\omega^2\overline{\tau}^2} \qquad (1)$$

where $\overline{\Delta}$ is the mean relaxation strength and $\overline{\tau}$ is the average relaxation time $\overline{\tau} = (\tau_\sigma \cdot \tau_\varepsilon)^{1/2}$ for a standard anelastic solid [32,33]. This is the classical Debye equation, which exhibits a maximum for the condition $\omega \cdot \overline{\tau} = 1$. The above expression (1) should apply for an internal friction spectrum obtained as a function of frequency in isothermal conditions. However, most times internal friction is measured as a function of temperature at a constant frequency and then the following expression should be applied to describe a relaxation peak [34], as we will do in next section:

$$Q^{-1} = \tan\phi = \frac{\overline{\Delta}}{2}\cdot\cosh^{-1}\left[\frac{E_a}{r_2\cdot k_B}\cdot\left(\frac{1}{T}-\frac{1}{T_P}\right)\right] \qquad (2)$$

being $E_a$ the activation enthalpy of the relaxation process, $T_P$ the peak temperature and $k_B$ the Boltzman constant. The expression includes the broadening factor $r_2$, which give account for the distribution broadening of the relaxation process, which is directly related to the β factor of the Gaussian distribution [32,34].

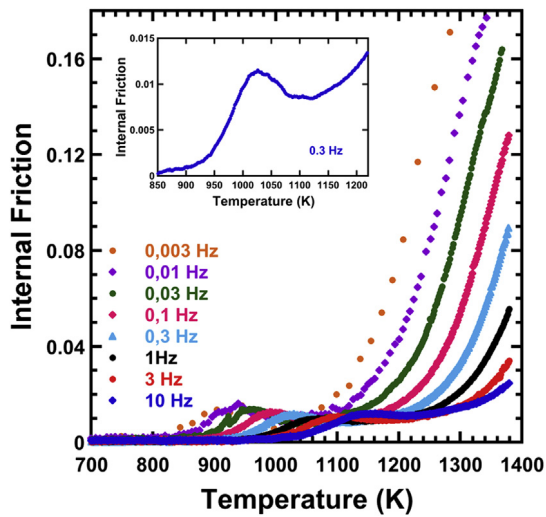

**Fig. 3.** Internal friction spectra measured at different frequencies for the as-cast material. A relaxation peak between 900 and 1100 K is observed, superimposed to a high temperature background, HTB.

The above expression (2) will be applied to analyze the relaxation peak appearing between 900 K and 1000 K, in Fig. 3, but the analysis of the HTB above 1100 K could be a bit more complex. On the one hand, the HTB could be considered as the low-temperature side of a relaxation peak with a maximum taking place at very high temperature, out of the measurement range. In this case the HTB is measured in an experimental condition verifying $\omega\cdot\overline{\tau} > 1$ and consequently the expression (1) will reduces to:

$$Q^{-1} = \tan\phi(\omega) = \overline{\Delta}\cdot\frac{1}{\omega\cdot\overline{\tau}} \qquad (3a)$$

On the other hand, at very high temperature and under the internal stresses, the material may not behave as a standard anelastic solid but rather than a Maxwell solid, as suggested by Weller et al. [24], undergoing like a micro-creep behaviour. The analysis of the Maxwell rheological model [24,33] predict a similar expression to (3a) but slightly simpler:

$$Q^{-1} = \tan\phi(\omega) = \frac{1}{\omega\cdot\overline{\tau}} \qquad (3b)$$

The difference between expressions (3a) and (3b) is directly related to the time constant of the process behaviour (standard or Maxwell), but for the short times involved during one oscillation in a dynamic experiment, both approaches cannot be experimentally distinguished. Moreover, Schoeck et al. [35] proposed a description of the HTB based on a generalized Maxwell rheological model including a distribution factor n that takes values from 0 up to 1 for the case of ideal viscoelasticity. Then the description of the internal friction during the HTB would be given by:

$$Q^{-1} = \tan \phi(\omega) = \frac{1}{(\omega \cdot \overline{\tau})^n} \qquad (4)$$

A similar expression was further used to establish the relationship between internal friction and creep in ceramic materials [36]. Since visco-elastic behaviour is determined by thermally-activated processes, an Arrhenius equation is expected for the relaxation time $\tau$:

$$\tau^* = \overline{\tau}^n = \tau_0^n \cdot \exp(n \cdot E_a / k_B \cdot T) \qquad (5)$$

where the distribution factor n play the same role than the broadening factor in expression (2) being $n \approx 1/r_2$ in a first approximation.

Then, in the next section we will use expression (2) to analyze the relaxation peak, and expressions (4) and (5) to analyze the HTB.

## 3. Results and analysis

Fig. 3 shows the internal friction (IF) versus temperature measured at different frequencies in the as-cast material. An internal friction peak is clearly observed between 900 and 1000 K, as shown in the insert for 0.3 Hz, superimposed to the high temperature background (HTB). Both the peak and the HTB are shifted towards lower temperatures when decreasing the frequency. In Fig. 4 the corresponding IF spectra measured at different frequencies on the extruded alloy, apparently show only a monotonous exponential increase of the HTB with no clear presence of any IF peak. However, when the spectra measured in both microstructural states are compared, as shown in Fig. 5, it can be observed that the HTB of the extruded sample is much higher than that for the as-cast sample, and the IF peak observed in the as-cast sample seems to be also present in the extruded sample, but is completely masked by the HTB. This difference in intensity of the HTB is at a first sight an indication of a lesser creep resistance (for the same temperature and stress) of the extruded sample in comparison with the as-cast sample, as confirmed by the creep measurements previously performed, see Fig. 8 in Ref. [29].

A quantitative analysis of the HTB can be conducted to determine the activation enthalpy of the mechanisms responsible for the anelastic deformation during the oscillation of the pendulum. Taking natural logarithms in Equation (4):

$$\ln(\tan \phi(T, \omega)) = -n \cdot \ln(\omega) - \ln(\tau^*(T)) \qquad (6)$$

Then the plot of $\ln(\tan\phi(T, \omega))$ versus $\ln(\omega)$ should give a straight line with a slope corresponding to the value of the distribution factor n. From the experimental results shown in Figs. 3 and 4, several temperatures have been selected and for each temperature $T_i$ the internal friction at each frequency has been measured. It is then possible to plot the logarithm of the internal friction versus logarithm of the corresponding frequency, Fig. 6, from the spectra of the as-cast alloy shown in Fig. 3. From the slopes of the linear regressions at each temperature $T_i$ it is possible to obtain the distribution factor n, see Equation (6), which will be the same for all temperatures. A value of n = 0.415 ± 0.005 is obtained for the as-cast alloy. The same analysis has been performed on the extruded alloy from the spectra of Fig. 4, and the results are plotted on Fig. 7 giving a distribution factor n = 0.264 ± 0.003.

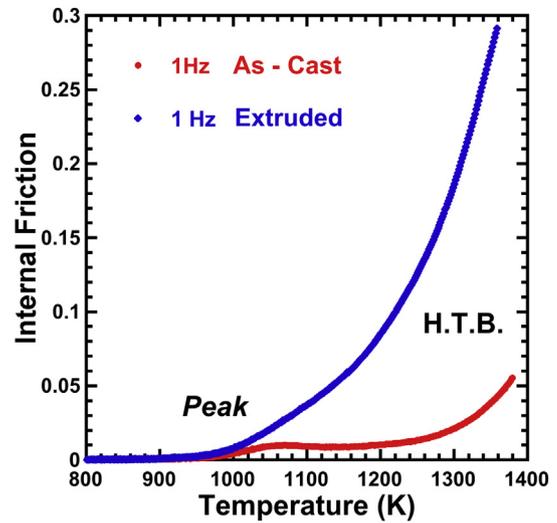

Fig. 5. Comparison of the internal friction spectra for both as-cast and extruded alloys.

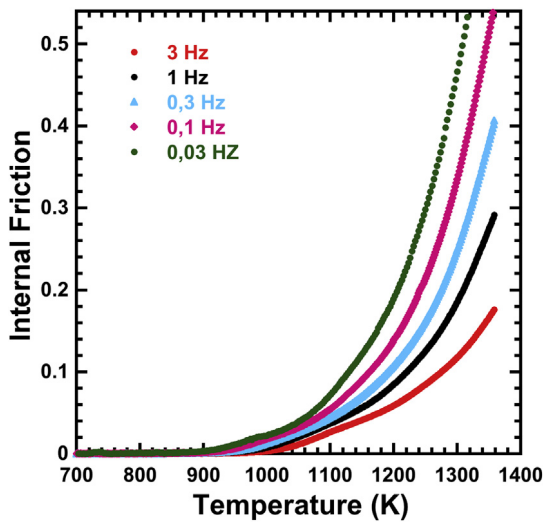

Fig. 4. Internal friction spectra measured at different frequencies for the extruded material. Apparently no relaxation peak is observed, but a monotonous exponential increase of the HTB in the internal friction at high temperature.

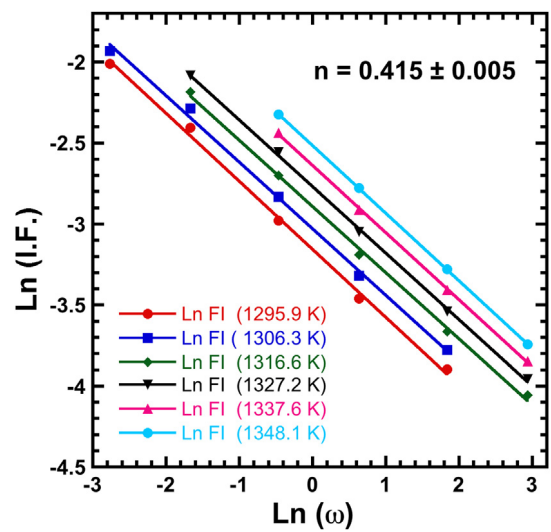

Fig. 6. $\ln(Q^{-1}) - \ln(\omega)$ at different temperatures, providing a value of the distribution factor of n = 0.415 for the as-cast material.

Moreover, the ordinate at the origin in expression (6), i.e. the intercept of $\ln(\tan\phi(T, \omega))$ with the axis $\ln(\omega) = 0$, provides the natural logarithm of the apparent relaxation time $\tau^*$ for each temperature. Considering the Arrhenius dependence of $\tau^*$ with temperature, Equation (5), we have:

$$\ln(\tau^*) = n \cdot \ln(\tau_0) + \frac{n \cdot E_a}{k_B \cdot T} \qquad (7)$$

and plotting $\ln(\tau^*) - 1/T$ it is possible to obtain the activation energy of the mechanism involved during deformation, from the slope of the straight line and taking into account the n factor previously determined. In addition, the ordinate at the origin allows determining the pre-exponential time $\tau_0$. Figs. 8 and 9 show $\ln(\tau^*) - 1/T$ plots for the as-cast and the extruded materials respectively. For the as-cast alloy is obtained an apparent activation energy $E_{cast}(IF) = 4.40 \pm 0.05$ eV, and a time $\tau_0 = 1.8 \times 10^{-14} \times 10^{\pm 1}$ s, whereas for the extruded alloy $E_{ext}(IF) = 4.75 \pm 0.05$ eV, and $\tau_0 = 2.8 \times 10^{-17} \times 10^{\pm 1}$ s. In addition, the values of the n factor allow calculation of the broadening factor $r_2$, and so the β distribution factor of the Gaussian distribution for the activation enthalpies: $r_2(cast) = 2.40 \Rightarrow \beta = 3.3$ and $r_2(ext) = 3.78 \Rightarrow \beta = 5.5$, according to the function described in Refs. [32,34]. These rather broad values correspond usually to dislocation processes and can be hardly associated to pure atomic diffusion processes by exchange with vacancies.

The activation energies obtained by IF experiments for the as cast and the extruded materials, 4.40 and 4.75 eV respectively, correlate well with those obtained from creep experiments, 4.1 and 4.35 eV [29]. The discussion of these values will be done in the next section.

Regarding the IF peak P1 appearing between 900 and 1000 K in Fig. 3 it should be pointed out that it clearly overlaps the HTB and consequently its analysis requires the subtraction of the HTB in order to isolate the relaxation peak P1. Now this is a relatively easy task because the HTB has been already analyzed and the obtained values of $E_a$, n, $\tau_0$ and $\omega$ can be used to calculate the function of HTB versus temperature through the expressions (4) and (5). To illustrate this fitting, the spectrum measured at 0.3 Hz in the as-cast alloy is plotted in Fig. 10, together with the HTB predicted by expressions (4) and (5). Then we may proceed to subtract the theoretical HTB from the experimental spectra in order to isolate the relaxation peak P1 also shown in Fig. 10. As the analysis of the IF relaxation peaks must be performed as a function of $1/T(K)$, this plot is also shown in the insert of Fig. 10. This subtraction and analysis have been performed for the spectra at different frequencies, and then the isolated relaxation peak P1 is plotted in Fig. 11a for five frequencies. First we have to notice the shift of the peak P1 towards low temperatures when decreasing the frequency, as corresponds to a relaxation peak. We have also to notice the shoulder on the high temperature side of the main peak P1, which can be named peak P2 because it also shifts whit frequency. In a first sight it is difficult to say if this second component P2 has some overlapping with the maximum of the main peak P1, preventing a correct determination of the shift maxima. Then, we have determined the activation enthalpy of P1 from the plot $\ln(\omega)$ versus $1/T$ for the points at half-eight of the low temperature side of the peak, which clearly is not affected by the shoulder component. The slope of the straight line from Fig. 11b give the activation energy associated to the relaxation peak P1, $E_a$(P1) = 2.99 eV ± 0.01 eV. Once we have measured the activation parameters of the peak P1, it is possible to obtain the theoretical relaxation peak using expression (2), which allows obtain the broadening factor $r_2 = 1.4$, for the best fitting shown in Fig. 12. This activation energy corresponds exactly to the Ti self-diffusion in $Ti_3Al$ ($\alpha_2$) phase [37,38], and the broadening factor corresponds to a point defect mechanism as well. Then, this relaxation peak P1 can be identified with the one reported in several TiAl intermetallics and attributed to the short distance diffusion of Ti atoms in $\alpha_2$ phase by exchange with a Ti-vacancy $V_{Ti}$ in between two Al atoms [27]. Under the oscillating stress this mechanism produces the reorientation of elastic dipoles of $Al-V_{Ti}-Al$, being responsible of a Zener-like relaxation process as recently described [27]. In what concerns the P2 component appearing as a shoulder in all the spectra of Fig. 11b, its analysis cannot be done in the frame of the present work, requiring further research. One important point to be remarked is that the relaxation peak P1 determines the temperature range from which the diffusion processes become thermally activated, indicating the onset of long distance diffusion processes controlling dislocation climbing mechanisms.

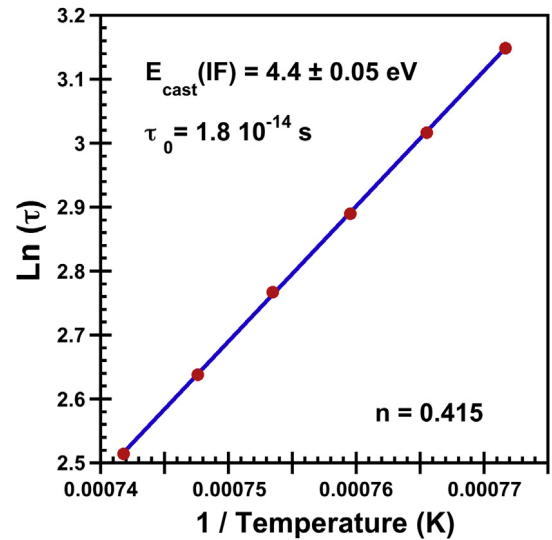

**Fig. 8.** Ln ($\tau^*$) versus 1/T for the as-cast material. From Equation (5), an activation energy of $E_{cast}(IF) = 4.40 \pm 0.05$ eV and a relaxation time $\tau_0 \cong 1.8 \times 10^{-14}$ s are obtained.

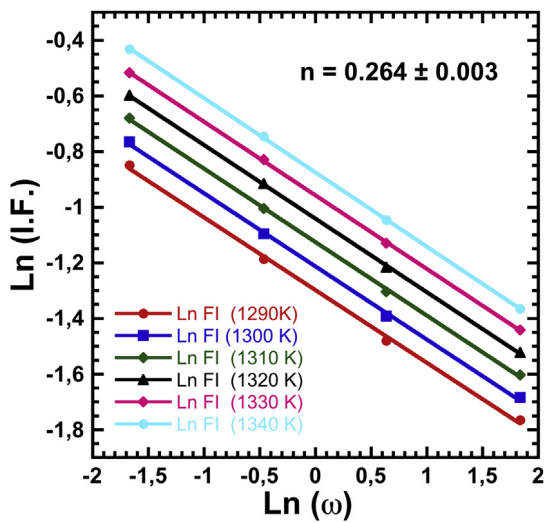

**Fig. 7.** $\ln(Q^{-1}) - \ln(\omega)$ at different temperatures, providing a value of the distribution factor of n = 0.264 for the extruded material.

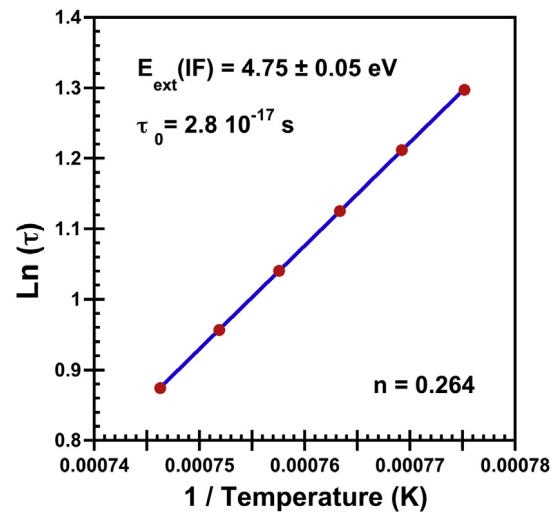

**Fig. 9.** Ln ($\tau^*$) versus 1/T for the extruded material. From Equation (5), an activation energy $E_{ext}(IF) = 4.75 \pm 0.05$ eV and and a relaxation time $\tau_0 \cong 2.8 \times 10^{-17}$ s are obtained.

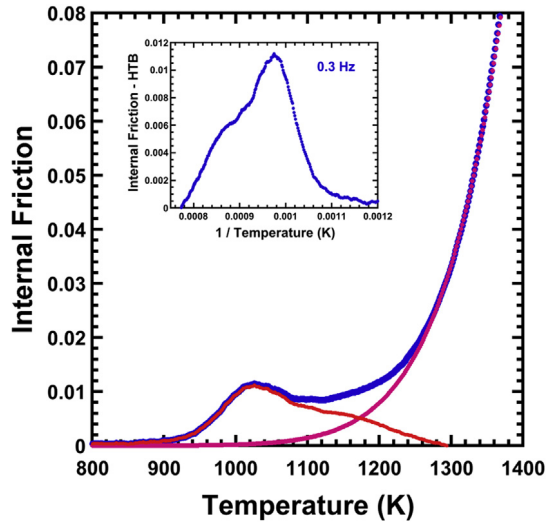

**Fig. 10.** Subtraction from the experimental results (blue dots) of the HTB calculated by expression (4) with the obtained parameters E, $\tau_0$ and n, (small magenta dots) for the particular case of the spectrum measured at 0.3 Hz in the as-cast alloy. The remaining peak (small red squares), has been also plotted as a function of 1/T in the insert.

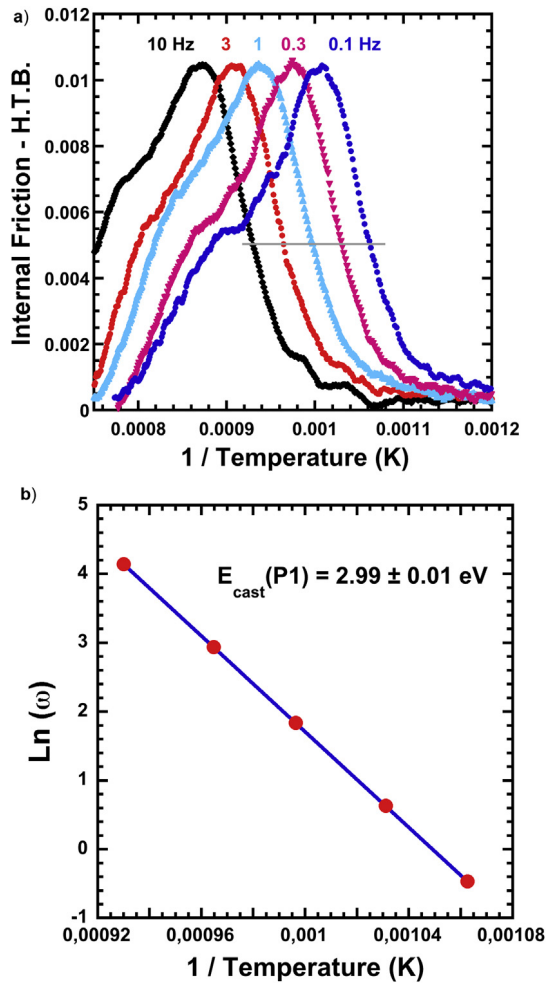

**Fig. 11.** (a) Relaxation peak P1 as a function of $T^{-1}$ for different frequencies, after subtraction the HTB from the experimental spectra. The shoulder component P2 is systematically evidenced as well. (b) Arrhenius diagram from the shift in temperature of the peak P1 at half height, allowing obtain the activation enthalpy E(P1) = 2.99 ± 0.01 eV.

## 4. Discussion

First we have to remark the excellent agreement between the internal friction HTB and creep measurements, which evidence that both techniques give very useful complementary information to reach a deep understanding of the microscopic mechanisms controlling the creep deformation at high temperature in intermetallics, like the γ-TiAl studied in the present work. However, in order to do a proper comparison between internal friction and creep measurements, a quantitative analysis of the results would be required. In particular we may ask about the microscopic mechanisms that sustain the experimental observations by both techniques above 1100 K.

A question arises about the determination of the activation energy for creep. It is well established that this determination is usually conducted in the steady state at each temperature where the creep substructures remain essentially unchanged with straining. Furthermore, a simple correlation has been established that correlates the creep data at various temperatures (originally in aluminium) by the use of a single value of the activation energy, namely the time-temperature parameter of the form $t \cdot \exp(-Q/RT)$ [39]. Since the activation energy for deformation is not a function of strain, even in the primary creep region, we have a justification for correlating the data at the various temperatures in order to determine the activation energy in the region just after the elastic regime that corresponds to the data obtained by the internal friction spectra.

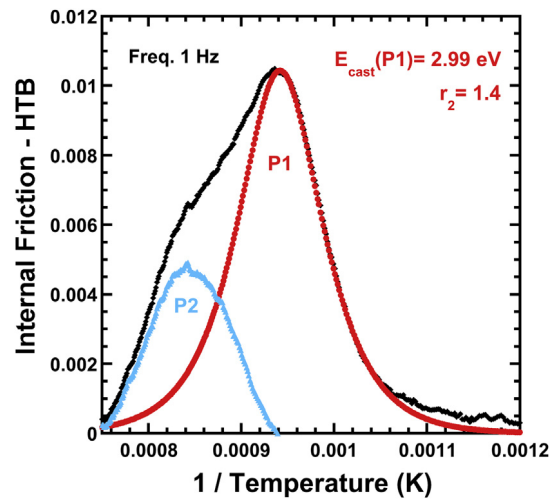

**Fig. 12.** De-convolution of the spectrum measured at 0.3 Hz, after subtraction the HTB. The experimental points (black diamonds) are fitted to a Debye peak P1 with the corresponding parameters (red dots). The subtraction of this peak allows isolate the shoulder component P2 (cyan triangles), which unfortunately cannot be properly fitted as to precisely obtain its activation parameters. (For interpretation of the references to colour in this figure legend, the reader is referred to the web version of this article.)

We may conclude that both techniques are seeing basically the same microscopic processes, although we have to note that creep measurements involve long distance mobility of atomic defects (dislocations, point defects etc) contributing to the plastic deformation, whereas internal friction act as a prove able to see only short distance processes corresponding to the first steps of the atomic defects motion. In other words, internal friction is seeing the motion of the defects at short distance during the activation of the mechanisms contributing to creep deformation.

To analyse the results obtained on the as-cast alloy, lets remember that in the original work from Jiménez et al. [29], creep and electron microscopy characterization allowed the conclusion that at 1173 K the as-cast alloy deforms by dislocation slip, in good agreement with the obtained values of the stress exponent at high temperature, near 4 and 5, corresponding to constitutive laws describing slip creep [40]. Then the authors consider that creep deformation must involve climb of dislocation and consequently the measured apparent activation energy for creep $E_{cast}(\text{creep}) = 4.1$ eV will be related to lattice diffusion of the slower specie [29]. In principle, two different defects mobility models can be considered to explain both creep and internal friction results.

### 4.1. Jog-dragging model

The mechanism of jog dragging was invoked as an additional source of glide resistance to ½ < 110] screw dislocations. Indeed, gliding of screw dislocations controlled by jog dragging, which climb by diffusion of vacancies, constitutes the "jog-dragging screw dislocation model" originally proposed by Barret and Nix [41] and later modified by Mills et al. [42,43] was considered as a rate-controlling creep behaviour. The jog-dragging model was also analyzed and proposed by Nó et al. [44,45] to explain an internal friction relaxation peak in ultra-pure aluminium after different deformation treatments including creep [46]. However, this model has some drawbacks because it does not allow explaining the strain rate during creep. In addition the expected activation enthalpy measured by IF must be the self-diffusion of the slowest specie on the corresponding lattice, this means 3.71 eV for Al in γ-TiAl phase [47,38], which is much lower than the 4.40 eV measured by IF in the present work for the as-cast material. Finally, the predicted activation volume is much higher [45] than the measured values, which lie between 15 and 60 b³, in this temperature range for similar γ-TiAl alloys [48]. So in principle this model does not allow to explain the whole set of experimental results by creep and IF and consequently can be disregarded.

### 4.2. Dislocation climb model

Climbing of dislocations was considered as the main controlling creep rate process in the regime with $n \leq 5$ [49] and many works since the nineteen's attributed the creep behaviour on γ-TiAl alloys, above 1000 K, to this mechanism [12,48,50]. This model predicts very well the creep strain rate and the stress exponent n. However, the expected activation enthalpy apparently should be the self-diffusion of the slowest specie, which as commented above is smaller than the experimentally measured values. The predicted activation volume, from 1 to 10 b³ [51–53], is also smaller than the experimentally observed. Nevertheless, we have to comment that such predicted activation parameters corresponds to simple climbing models and a bit more complex model should be considered, as suggested Hirth & Lothe [54] to explain climb controlled creep. In addition, up to now there is no any model proposed to justify the internal friction associated to a dislocation climb mechanism. The discussion of these two points will be approached in the following paragraph.

### 4.3. Dislocation climb by jog-pair formation model

Let us consider the climbing process by vacancy diffusion towards the dislocation line and the jog-pair formation (JPF) mechanism as responsible for the global dislocation climb. In this case the activation energy of the climbing process should not be just the one for self-diffusion $E_{sd}$, but it must include the term associated to the required energy for the formation of a jog-pair $E_{jp}$, as predicted by Hirth & Lothe [54] and further also considered by Caillard & Martin [55]. Vacancy diffusion can take place through the lattice, involving the self-diffusion energy $E_{sd}$, or eventually by pipe diffusion. However, pipe diffusion mechanisms are invoked whenever the measured activation energy is lower than the one for self-diffusion, which is not the present case. This fact, together with the observed compact core of the ½ < 110] dislocations in γ-TiAl and its atomic ordered line [14] are reasons good enough to justify why pipe diffusion has not been considered in climb processes in γ-TiAl and it will be disregarded in the following analysis. Then in the present case we have to consider the self-diffusion coefficient:

$$D_{sd} = a^2 \cdot \nu_D \cdot \exp\left(\frac{-E_{sd}}{k_B \cdot T}\right) \tag{8}$$

where $E_{sd}$ is the self-diffusion energy of the involved atomic specie, $a$ the average interatomic distance and $\nu_D$ the Debye frequency. Then according the expression from Refs. [54,55] when climbing of dislocation is taking place by the jog-pair formation mechanism, the dislocation climb velocity during creep is given by:

$$\nu = 4\pi \cdot a \cdot \nu_D \cdot \frac{\tau \cdot \Omega}{k_B \cdot T} \cdot \exp\left[\frac{-\left(E_{sd} + \frac{1}{2} E_{jp}\right)}{k_B \cdot T}\right] \tag{9}$$

where $\Omega$ is the atomic volume $\tau$ the resolved climbing stress per unit of length. This mechanism, which was developed to explain climb controlled creep, can be also applied to explain the internal friction at high temperature. Nevertheless, the development of the internal friction model, which will be fully described elsewhere [56], is out of the scope of the present paper and we will focus here on the theoretical expression of the activation parameters for creep and internal friction, in order to compare such predictions with the experimentally obtained values.

During the jog-pair formation process, which is schematically described in Fig. 13, vacancies will diffuse towards the dislocation line, Fig. 13a, given place to the formation of a pair of jogs, Fig. 13b, which will move laterally to allow dislocation climb, Fig. 13c. However we have to consider the jog–jog interaction in a similar way than in the kink-pair formation model proposed by Seeger [57] and reviewed in Refs. [54,55,58]. In this model the energy for jog-pair formation when the jogs are separated the critical distance $d_c$ at the saddle point is given by:

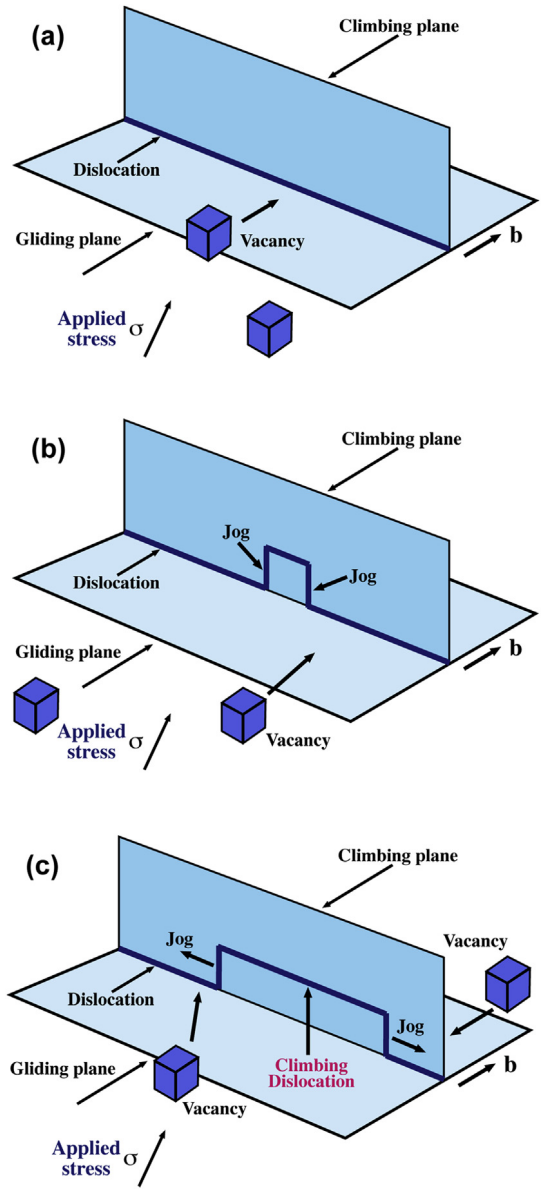

**Fig. 13.** Schematic description of the dislocation climb mechanism by vacancy diffusion and jog-pair formation. (a) A vacancy diffuses through the lattice towards the dislocation line. (b) A jog-pair is formed on the dislocation line. (c) When new vacancies diffuse towards the dislocation line, the jogs migrate laterally allowing the climbing of the dislocation.

$$E_{jp}(d_c) = 2E_j - \frac{\mu \cdot h^2 \cdot b^2}{8\pi(1-\nu) \cdot d_c} - (h \cdot b \cdot d_c) \cdot \sigma \tag{10}$$

where $E_j$ is the self-energy of a single jog of height h, which for an edge dislocation is given by Ref. [54]:

$$E_j = \frac{\mu \cdot h \cdot b^2}{4\pi(1-\nu)} = \frac{\mu \cdot b^3}{4.9\pi(1-\nu)} \tag{11}$$

The second term in (10) corresponds to the jog–jog interaction and the third term is the work of the applied stress. The critical distance dc at the saddle point is determined by:

$$d_c = \left(\frac{h \cdot b}{8\pi(1-\nu)} \cdot \frac{\mu}{\sigma}\right) \tag{12}$$

In this model, the dislocation climb velocity is given by (9) provide that the mean free path λ of a jog issued from a pair [54,55] remain smaller than the dislocation length L, so the condition λ < L should be fulfilled, with λ:

$$\lambda = a \cdot \exp\left(\frac{E_{jp}}{2k_B \cdot T}\right) \tag{13}$$

According expression (9), the model of dislocation climbing by vacancy diffusion through the lattice and jog-pair formation predicts an apparent activation energy:

$$E_a = E_{sd} + \tfrac{1}{2} E_{jp} \tag{14}$$

To do a quantitative estimation of the predicted values for creep and internal friction, and to compare these values with the experimental ones, we have to consider the real values of all input parameters for γ-TiAl. In Table S1 of the Supplementary material all the required data of modulus and lattice parameters from the literature have been included for the TiAl phases, and in most of cases we have calculated the corresponding values at 1173 K. We have chosen this temperature, as common reference for all comparisons, because it corresponds (900 °C) to the mean value in the high temperature range for creep measurements [29], and it also represents the onset of the HTB analysed in internal friction measurements. So all parameters, in expressions (8)–(14), required to test the JPF model are corrected at 1173 K.

In internal friction measurements the maximum applied stress has always been $\sigma = 10^{-5}\mu$, while in creep measurements there were a slight variation on the stress value used to calculate the activation energy, $\sigma = 9 \cdot 10^{-4} E$, for the as-cast alloy and $\sigma = 10^{-4}E$, for the extruded alloy [29]. So the mean value will be considered, which according the ratio of the Young and shear modulus is: $\sigma = 5 \cdot 10^{-4}E = 1.233 \cdot 10^{-3}\mu$.

With the above data, the self-energy of a single jog in a ½ < 110] edge dislocation in γ-TiAl is obtained from (11), $E_j = 0.701$ eV. The critical distance $d_c$ from (12) allows to obtain a qualitative description of the activation volume $v_a = hbd_c$ for the JPF mechanism, which depend on the stress through the term $d_c$. At this point it is interesting to remark that the model of dislocation climbing by JPF predicts an activation volume between 10 and 65 $b^3$, not so small as traditionally proposed for dislocation climb (between 1 and 10$b^3$) [51–53], and in rather good agreement with the experimental values measured at high temperatures, between 15 and 70 $b^3$, in similar γ-TiAl alloys [12,48,59]. To evaluate the work of the stress in expression (10) we have used the value of $v_a = 40$ $b^3$ measured by stress relaxation in a single phase γ-TiAl alloy [48] at 1173 K, which is our reference temperature, and a good mean value from the literature results. With these parameters we can use (10) to calculate the JPF energy Ejp in both conditions: during internal friction, $E_{jp}(IF) = 1.394$ eV, and during creep, $E_{jp}(creep) = 0.937$ eV, measurements. These data have been reported in Table 1 for comparison with the experimental results. From the obtained values for Ejp we can verify, using (13), if the condition λ < L is actually fulfilled; indeed it is because λ values are 30 nm for creep conditions and 145 nm for internal friction, which in both cases are smaller than the length of dislocations observed by electron microscopy in these samples [29] and in similar γ-TiAl alloys [14].

Finally the apparent activation energy predicted by the model is obtained from (14) and compared with the experimental values measured by internal friction and creep. In the case of the as-cast alloy we may consider that the process is controlled by the diffusion of Al atoms in γ-phase, as suggested in [29], which is $E_{sd}$(Al-γ) = 3.71 eV [47,38]. Then the theoretical activation energies predicted by the model will be $E_{cast}$(IF) = 4.407 eV and $E_{cast}$(creep) = 4.178 eV, in excellent agreement with both experimental results, as shown in Table 1.

**Table 1**
Predicted values for the different terms of the dislocation climb by vacancy diffusion and jog-pair formation (JPF) model, for both internal friction and creep measurements, and comparison with the experimentally measured values. The results for both microstructural states of the alloy, as-cast and extruded, are presented.

| | | $E_j$ (eV) | $d_c$ (b unit) | $E_{jj}$ (eV) | $v_a\sigma$ (eV) | $E_{jp}$ (eV) | $\frac{1}{2}E_{jp}$ (eV) | $E_{sd}$ (eV) | $E_a$ (eV) (model) | $E_a$ (eV) (experimental) |
|---|---|---|---|---|---|---|---|---|---|---|
| As-Cast | Int. friction | 0.701 | 65 | 0.005 | 0.003 | 1.394 | 0.697 | 3.71 | 4.407 | 4.40 |
| | Creep | 0.701 | 6 | 0.059 | 0.406 | 0.937 | 0.468 | 3.71 | 4.178 | 4.1 |
| Extruded | Int. friction | 0.701 | 65 | 0.005 | 0.003 | 1.394 | 0.697 | 4.08 | 4.777 | 4.75 |
| | Creep | 0.701 | 6 | 0.059 | 0.406 | 0.937 | 0.468 | 4.08 | 4.548 | 4.35 |

Let consider now the case of the extruded alloy. The microstructure of the extruded alloy is constituted by small grains, and creep behaviour is expected to be controlled by grain boundary sliding GBS (at low $\sigma$) as it is usually associated to a stress exponent close to 2 (in our case n = 2.2 [29]). However the measured activation energies measured by creep, $E_{ext}$(creep) = 4.35, as well as by internal friction, $E_{ext}$(IF) = 4.75 eV, are too high to be justified just by a pure atomic diffusion process. So, we may wonder about the particular mechanism that could be activated at grain boundaries to promote grain boundary sliding. From a recent overview on creep in metals [40] we may say that at present it is accepted that GBS involves the movement of dislocations even during superplastic behaviour. Ruano and Sherby [49] formulated phenomenological equations with n = 2 (close to the stress exponent in the present alloys n = 2.2) which corresponds to the GBS accommodated by dislocations movement taking place either within the grains or along grain boundaries. Taking into account the high density of dislocations reported in the literature [12,14] along the interfaces of the $\gamma/\alpha_2$ lamellae we may assume that an intensive dislocation motion of dislocations will take place in the $\gamma$ lamellae along these interfaces, to give account of the higher strain rate observed in the extruded material in comparison with the as-cast material [29]. This intensive motion of defects is also observed during internal friction measurements given place to a HTB much higher in the extruded alloy than in the as-cast alloy, as was shown in Fig. 5. At this point we may assume that similar mechanism should operate in both alloys (as-cast and extruded) and the intensive dislocation climbing by diffusion and jog-pair formation mechanism will be exhausted at the local scale along the grain boundary because the depletion of vacancies in the $\gamma$-lattice. To restore the local equilibrium vacancy concentration and to maintain the mechanism going on, vacancies must migrate from the surrounding environment, so also from the $\alpha_2$-lattice at the other side of the interface and again both atomic species Ti and Al must diffuse, the slower one controlling the process, this means the Al. This is a reasonable scenario if we consider that in the extruded microstructure, Fig. 2, the $\gamma$ phase grains are in general surrounded by $\alpha_2$ grains. In this case, the activation enthalpy for Al diffusion in $\alpha_2$ phase is $E_{sd}$(Al-$\alpha_2$) = 4.08 eV [37,38], and consequently should be the one controlling the process. Then, introducing this diffusion energy in (14), the theoretical activation energy predicted by the model can be obtained for both internal friction, $E_{ext}$(IF) = 4.777 eV, and creep, $E_{ext}$(creep) = 4.548 eV, conditions, as reported in Table 1. It is worthy of remark the good agreement that is also observed in this case, although the theoretical values are slightly higher than the experimental ones. This is not surprising because the diffusion of Al in $\alpha_2$ phase (4.08 eV) would take place only locally at some interfaces, contributing to the highest value of the activation energy, while in other interfaces the mechanism will not be exhausted and the diffusion of Al in $\gamma$ phase (3.71 eV) would be controlling the process determining the lowest value of the activation energy. In a realistic scenario it could be expected that a distribution between these two limits would locally operate at the interfaces given place to an intermediate value of the apparent activation energy, as experimentally observed. This description allows also to understand the broad distribution factor β = 5.5 for the activation energy measured by internal friction in the case of the extruded alloy.

## 5. Conclusions

Mechanical spectroscopy tests have been performed in a $\gamma$-TiAl intermetallic of Ti−46.8Al−1Mo−0.2Si (in at.%), in two different micro-structural conditions: as-cast and extruded. The high temperature behaviour of both alloys have been approached by the analysis of the internal friction spectra as a function of temperature, and the results compared with those obtained by creep tests in a previous work on the same alloys [29]. At the light of the presented results and their analysis, the following conclusions can be established:

* Internal friction spectra show a relaxation peak P1, at about 1050 K (at 1 Hz), with an activation energy of E(P1) = 2.99 ± 0.01 eV, which has been attributed to stress-induced reorientation of elastic dipoles Al−$V_{Ti}$−Al in $\alpha_2$-Ti$_3$Al phase by diffusion of a Ti atom, as previously reported in Ref. [27].

* The high temperature background (HTB) of the internal friction spectra has been used to characterize the microscopic mechanisms operating at high temperature. The characteristic parameters controlling the HTB in the as-cast alloy have been determined, $E_{cast}$(IF) = 4.40 ± 0.05 eV, $\tau_0$ = 1.8 × 10$^{-14}$ (×10$^{±1}$) and β = 3.3, whereas for the extruded alloy $E_{ext}$(IF) = 4.75 ± 0.05 eV, $\tau_0$ = 2.8 × 10$^{-17}$ (×10$^{±1}$) and β = 5.5, have been found.

* These values of the activation enthalpies measured for the internal friction HTB correlate very well with those previously measured by creep tests [29], $E_{cast}$(creep) = 4.1 eV and $E_{ext}$(creep) = 4.35 eV. The slight difference is justified by the difference of the applied stresses involved in both kind of measurements, internal friction and creep tests. This is an outstanding result because, independently of the proposed mechanisms, evidences that internal friction offers reliable quantitative data about the parameters controlling high temperature creep.

* The mechanism based on dislocation climbing by self-diffusion and jog-pair formation has been proposed to explain both the creep and internal friction measurements. This model has been quantitatively evaluated taking into account that the mechanism should be controlled by the slower diffusion atomic species involved during the process. In the as-cast alloy the process is controlled by Al atom diffusion in $\gamma$-TiAl, whereas in the extruded alloy the creep deformation by GBS requires the participation of both phases across the interface and the process

would be locally controlled by Al atom diffusion in γ-TiAl phase as well as in α₂-Ti₃Al phase, given place to a broad activation energy distribution, as experimentally measured.

* The predictions of the model exhibit an exceptionally good agreement between the theoretically expected values and the experimental ones, as shown in Table 1. In addition the model also explains the slight differences found between internal friction and creep measurements.

At the light of the present work we may also conclude that internal friction measurements, through the HTB analysis, is a very useful technique to study the mechanisms controlling creep deformation at high temperature. The obtained data are reliable and comparable to those obtained by creep and the use of internal friction in a collaborative way with creep tests could be of a great scientific and technological interest.

### Acknowledgements

This work was supported by the Spanish MICINN project CONSOLIDER-INGENIO 2010 CSD2009-00013, as well as by the Consolidated Research Group IT-10-310 from the Education Department and the project ETORTEK ACTIMAT from the Industry Department of the Basque Government.

### Appendix A. Supplementary data

Supplementary data related to this article can be found at http://dx.doi.org/10.1016/j.actamat.2015.09.052.